\documentclass[prd,aps,preprint,epsfig,floats]{revtex4}
\usepackage{graphics}
\usepackage{epsfig}

\begin{document}

\title{\Large\bf An SO(10) grand unification model with S4 flavor symmetry}
\author{\bf Yi Cai and Hai-Bo Yu }

\affiliation{ Department of Physics, University of Maryland, College
Park, MD 20742, USA}
\date{\today}

\begin{abstract}

We present a supersymmetric grand unification model based on SO(10)
group with $S4$ flavor symmetry. In this model, the fermion masses
are from Yukawa couplings involving $\bf{10}$ and
$\overline{\bf{126}}$ Higgs multiplets and the flavor structures of
mass matrices of both quarks and leptons are determined by
spontaneously broken $S4$. This model fits all of the masses and
mixing angles of the quarks and leptons. For the most general
CP-violation scenario, this model gives $\sin\theta_{13}$ a wide
range of values from zero to the current bound with the most
probable values $0.02-0.09$. With certain assumptions where leptonic
phases have same CP-violation source as CKM phase, one gets a
narrower range $0.03-0.09$ for $\sin\theta_{13}$ with the most
probable values $0.04-0.08$. This model gives leptonic Dirac CP
phase the most probable values $2-4$ radians in the general
CP-violation case.
\end{abstract}

\maketitle
\section{Introduction}

The discovery of no-zero neutrino masses and lepton mixings have
raised hope to understand the mystery of flavor structures of quarks
and leptons in a unified way\cite{rev}. Although many similarities
between leptons and quarks make such unification plausible, the
mixing pattern in the lepton sector is very different from that in
the quark sector. However, there are now many grand unification
models based on $SO(10)$ gauge group that can give small quark
mixings and large lepton mixings along with all their masses with
few assumptions\cite{rev}.

Another interesting possibility is that there may exist horizontal
underlying flavor symmetry. This is favored by leptonic mixing
pattern with near maximal atmospherical mixing angle and the
vanishing $\theta_{13}.$ A permutation symmetry between $\mu$
neutrino and $\tau$ neutrino in the flavor basis has been proposed
in recent years\cite{mutau}\cite{moh}. Even though there is no
apparent evidence of such symmetry in charged lepton and quark
sector, it has been shown that the unified description of quarks and
leptons with this symmetry is possible\cite{mutaugut}. Applications
of higher permutation group $S3$, $S4$ and $A4$ to flavor symmetry
also have been discussed in the
literature\cite{s3}\cite{s4}\cite{a4}. Other discrete groups such as
dihedral group $D4$ and $D5$ have been studied \cite{d4}\cite{d5}.

In this paper, we focus on the group $S4\times SO(10)$. $S4$ has
certain good features to be a flavor symmetry. First, it has three
dimensional irreducible representation to accommodate the three
generations of fermions naturally. Note that this is different
from $S3$ because the largest irreducible representation of $S3$
has dimension two and therefore we have to treat one family of
fermions different from other two. Second, it can be embedded into
continuous group $SU(3)$ or $SO(3)$\cite{s4HLM}. As we will show
below, $S4$ symmetry also gives degenerate spectrum of the
right-handed neutrinos naturally, which has some interesting
consequences for the neutrino phenomenology. For example, in this
case, one can use the resonant enhancement of leptogenesis for
(quasi-)degenerate right-handed neutrinos to generate enough
baryon asymmetry\cite{lepto}. With the degenerate heavy
right-handed neutrinos, the low energy neutrino flavor structure
is determined by Dirac mass matrix at the seesaw scale completely,
which makes it easier to reconstruct high energy physics from low
energy observables. Some work has been done in this direction. In
Ref.\cite{s4LM}, Lee and Mohapatra constructed a $S4\times SO(10)$
model, which naturally gives quasi-degenerate spectrum of
neutrinos masses with small solar angle, which already has been
ruled out by large mixing angle MSW solution to the solar neutrino
problem. In principle radioactive corrections may amplify the
solar angle and keep the other two angles unchanged, but generally
this needs extreme fine-tuning of parameters at the seesaw scale
to realize it. On the other hand, in a recent paper\cite{s4HLM} by
Hagedorn, Linder, and Mohapatra, a low energy scale
non-supersymmetric model is presented based on $S4$ flavor
symmetry, which can accommodate current neutrino data. Our goal is
to see if we can embed the model of Ref.\cite{s4HLM} into a SUSY
GUT framework without running into the small solar angle problem
of Ref.{\cite{s4LM}}. In this letter, we address this question and
find that we can build a realistic model based on $S4\times
SO(10)$ with the proper choice of the parameter space.

In this model, all the quarks and leptons of one generation are
unified into a $\bf{16}$ spinor representation of $SO(10)$ and the
Yukawa coupling structures of three generations are determined by
$S4$. We use ${\bf 10}$ and ${\bf\overline {126}}$ representations
of $SO(10)$ for Yukawa couplings to account for all the fermions
masses and mixing angles\cite{minso10}\cite{others}. Even though
in the most general CP-violation case this model has $18$ complex
parameters, it is not obvious whether it can accommodate all
observed masses and mixing angles because of constraints from $S4$
flavor symmetry and the correlations between quarks and leptons
indicated by $SO(10)$ unification. For instance, with the particle
assignment of $S4$ in this model, the heavy right-handed neutrino
mass matrix is proportional to an identity matrix, and the Dirac
mass matrix of neutrino determines the mixing among light
neutrinos completely. The general mechanism to generate the lepton
sector mixing independently from the quark sector by right-handed
neutrinos does not work in this model. On the other hand, one may
argue that since the total number of parameters is much larger
than that of obervables, this model may lose predicability even if
it can fit all the obervables. We find this not to be the case. It
turns out that half of complex phases can be rotated away by
choices of basis and redefinitions of the right-handed fields of
charged leptons and down-type quarks. For the most general
CP-violation case, this model gives wide range of
$\sin\theta_{13}$ from zero to current bound with the most
probable values $0.02-0.09$. The most probable values of leptonic
CP phase are $2-4$ radians. With certain assumptions where the
leptonic phases have same CP-violation source as CKM phase, one
gets narrower predicted range $0.03-0.09$ for $\sin\theta_{13}$
with the most probable values $0.04-0.08$.

Some issues about Higgs sector still need to be addressed. As we
have six ${\bf 10}$s and three ${\bf\overline{ 126}}$s, without
analyzing the $S4\times SO(10)$ invariant Higgs potential, whether
or not we can get the desired vacuum configuration still remains an
open question. We do not concern with doublet-doublet splitting and
doublet-triplet splitting problems in this paper. With such rich
Higgs fields, we assume they can be realized in some way. And
another fact we should be careful is that generally the discrete
flavor symmetry can enhance the accidental global symmetry of Higgs
potential and lead to unwanted massless Nambu-Goldstone bosons.
There are ways found in the literature to avoid it. One can
introduce gauge singlet Higgs fields whose couplings are invariant
under discrete symmetry but break the global symmetry\cite{kubo1},
or introduce soft terms which break discrete symmetry and global
symmetry\cite{kubo2}.

The paper is organized as follows: in Section 2, we present an
$SO(10)$ model with $S4$ flavor symmetry and present the mass
matrices of quarks and leptons; in Section 3, we present a detailed
numerical analysis including CP violation in quark and lepton
sector. We end with conclusions and remarks in Section 4.

\section{SUSY SO(10) Model with S4 flavor symmetry}
The group $S4$ is the permutation group of the four distinct
objects, which has 24 distinct elements. It has five conjugate
classes and contains five irreducible representations
$\bf{1}$,$\bf{1'}$,$\bf{2}$,$\bf{3}$ and $\bf{3'}$. Our assignment
of fermions and Higgs multiplets to $S4\times SO(10)$ are shown in
Table \ref{S4charge}.

\begin{table}[h]
\begin{center}
\begin{tabular}{|c| c| c| c| c| c| c|} \hline \hline
Fermions & \multicolumn{6}{|c|}{Higges Bosons}\\
\hline $\Psi_a$,$a$=1,2,3
&$\Phi$&$\overline\Delta_0$&$\overline\Delta_{1,2}$
&$H_0$&$H_{1,2}$&$H_{3,4,5}$\\
\hline $\{\bf{3'}\}\times\{\bf{16}\}$&$\{{\bf 1}\}\times\{{\bf
210}\}$&$\{\bf{1}\}\times\{{\bf\overline{126}}\}$&$\{\bf{2}\}\times\{\bf{\overline{126}}\}$&
$\{\bf{1}\}\times\{\bf{10}\}$&$\{\bf{2}\}\times\{\bf{10}\}$&$\{\bf{3}\}\times\{\bf{10}\}$\\
\hline \hline
\end{tabular}
\end{center}
\caption{Transformation property of fermions and Higgs multiplets
under $S4\times SO(10)$} \label{S4charge}
\end{table}
In this model, we assign three generations of $\bf{16}$ to $\bf{3'}$
irreducible representation of S4, because $\bf{3'}$ can be
identified with the fundamental representation of continuous group
SO(3) or SU(3)\cite{s4HLM}\footnote{If one gives up the possible
embedding of $S4$ group to continuous group, one can choose $\bf{3}$
and the mass matrices for fermions do not change.}. In Higgs sector,
because of $\bf{3'}\times\bf{3'}=\bf{1}+\bf{2}+\bf{3}+\bf{3'}$, to
make Yukawa coupling $S4$ invariant, Higgs fields can not belong to
$\bf 1'$. $\bf 1$ is necessary for phenomenological reason,
otherwise all of the mass matrices would be traceless. To get
symmetric mass matrices which is required by group structure of
${\bf 16\cdot16\cdot10}$ or ${\bf 16\cdot16\cdot\overline{126}}$,
Higgs should not belong to $\bf {3'}$. We include both $\bf {2}$ and
$\bf {3}$ to get realistic mass and mixing of quark and lepton. One
might think six $\bf {10}$ Higgs fields transforming as $\bf {1}+\bf
{2}+\bf {3}$ under $S4$ are enough. But there are two reasons why we
also need ${\bf\overline{ 126}}$, one is to give right-handed
neutrinos heavy masses and the other is to fix the bad mass relation
between quark sector and lepton sector indicated by ${\bf
16\cdot16\cdot10}$. In this sense, our choice of Higgs fields is
minimal.

The breaking of $SO(10)$ to Standard Model(SM) can be realized in
many ways. In this model, we choose ${\bf 210}$ Higgs field, which
is ${\bf 1}$ under $S4$ transformation, to break $SO(10)$ to
$SU(2)_L\times SU(2)_R\times SU(4)_C$ $(G_{224})$ while keep the
$S4$ symmetry. We choose $({\bf 1},{\bf 3},{\bf 10})$ components of
only $\overline\Delta_0$ (the numbers denote representation under
the $G_{224}$) to get vev $v_R$ that breaks $G_{224}$ down to the SM
and gives heavy masses to right-handed neutrinos. With this breaking
pattern, $S4$ symmetry is kept down to the electroweak scale.

To see what this model implies for fermion masses, let us first
explain how the MSSM doublets emerge. Besides the $SU(2)_L$ Higgs
doublets from submultimplets $\bf (2,2,1)$ and $\bf (2,2,15)$
contained in ${\bf 10}$ and ${\bf \overline{126}}$ respectively,
we also have Higgs doublets contained in ${\bf (2,2,10)}\oplus{\bf
(2,2,\overline {10})}$ from ${\bf 210}$. Furthermore, to obtain
anomaly-free theory, we need to introduce three ${\bf 126}$, which
we denote by $\Delta$, that also contain Higgs doublets.
Altogether, we have fourteen pairs of Higgs doublets:
$\phi_u=(H_{iu},\overline{\Delta}_{ju},\Delta_{ju},\Phi_{u1},\Phi_{u2})$,
$\phi_d=(H_{id},\overline{\Delta}_{jd},\Delta_{jd},\Phi_{d1},\Phi_{d2})$,
where $i=0,...,5$ and $j=0,...2$. As noted, six pairs from $H$s,
three pairs from ${\overline{\Delta}}$s, three pairs from
${{\Delta}}$s and two pairs from ${\Phi}$. We can write Higgs
doublet mass matrix as $\phi_u M_H\phi_d^T$. $M_H$ can be
diagonalized by $XM_HY^T$, which $X$ and $Y$ are unitarity
matrices acting on $\phi_u$ and $\phi_d$ respectively. At the GUT
scale, by some doublet-triplet and doublet-doublet splitting
mechanisms, we assume only one pair of linear combinations of
$X^*_{\alpha\beta}\phi_{u\beta}$ and
$Y^*_{\alpha\beta}\phi_{d\beta}$, say $X^*_{1\beta}\phi_{u\beta}$
and $Y^*_{1\beta}\phi_{d\beta}$, has masses of order of the weak
scale and all others are kept super heavy near GUT scale, which
generally can be realized by one fine-tuning of the parameters in
the Higgs mass matrix. The MSSM Higgs doublets are given by this
lightest pair: $H^{\rm MSSM}_u=X^*_{1\beta}\phi_{u\beta}$ and
$H^{\rm MSSM}_d=Y^*_{1\beta}\phi_{d\beta}$. Since we focus on the
structures of Yukawa couplings, we do not discuss the details of
the splitting mechanisms that lead to the above results.

With Higgs fields and fermions listed in Table\ref{S4charge}, we
can write down $S4\times SO(10)$ invariant Yukawa coupling as
\footnote{For the products and Clebsch-Gordan coefficients of $S4$
group, one can see for example the Appendix A of
Ref.\cite{s4HLM}.}

\begin{eqnarray}
\nonumber W_{\rm
Yukawa}&=&(\Psi_1\Psi_1+\Psi_2\Psi_2+\Psi_3\Psi_3)(h_0
H_0+f_0\bar\Delta_0)\\\nonumber &&+  \frac{1}{\sqrt
2}(\Psi_2\Psi_2-\Psi_3\Psi_3)(h_1H_1+f_2\bar\Delta_1)+\frac{1}{\sqrt
6}(-2\Psi_1\Psi_1+\Psi_2\Psi_2+\Psi_3\Psi_3)(h_1H_2+f_2\bar\Delta_2)
\\&&+
h_3[(\Psi_2\Psi_3+\Psi_3\Psi_2)H_3+(\Psi_1\Psi_3+\Psi_3\Psi_1)H_4+(\Psi_1\Psi_2+\Psi_2\Psi_1)H_5].
\end{eqnarray}

After electroweak symmetry breaking, $({\bf 2,2,1})$ of
$H_i$$(i=0,...,5)$ component acquires vevs (denoted by $\langle
H_i\rangle^u$ and $\langle H_i\rangle^d$). And $({\bf 2,2,15})$
sub-multiplet of
$\overline\Delta_j$$(j=0,...,2)$ also get induced vevs. 
Their vevs are denoted by $\langle\overline\Delta_j\rangle^u$ and
$\langle\overline\Delta_j\rangle^d(j=0,1,2)$.

The mass matrices for the quarks and the leptons have following sum
rules:
\begin{eqnarray}
M_u&=&M_u^{(10)}+M_u^{(126)}\label{upmass1},\\
M_d&=&M_d^{(10)}+M_d^{(126)}\label{dmass1},\\
M^D_\nu&=&M_u^{(10)}-3M_u^{(126)},\\
M_l&=&M_d^{(10)}-3M_d^{(126)}\label{lmass1},\\
M_\nu&=&-{M^D_\nu}^T M^D_\nu/f_0v_R,
\end{eqnarray}
where
\begin{eqnarray}
M_u^{(10)}&=& \left[\matrix{a_0-2a_2&a_5&a_4\cr
a_5&a_0+a_1+a_2&a_3\cr a_4&a_3&a_0-a_1+a_2}\right],\\
M_d^{(10)}&=&\left[\matrix{b_0-2b_2&b_5&b_4\cr
b_5&b_0+b_1+b_2&b_3\cr b_4&b_3&b_0-b_1+b_2}\right],\\
M_u^{(126)}&=&\left[\matrix{d_0-2d_2&0&0\cr 0&d_0+d_1+d_2&0\cr
0&0&d_0-d_1+d_2}\right],\\
M_d^{(126)}&=&\left[\matrix{e_0-2e_2&0&0\cr 0&e_0+e_1+e_2&0\cr
0&0&e_0-e_1+e_2}\right],
\end{eqnarray}
and where $a_i$ and $b_i$ are products of the type $h\langle
H_i\rangle^u$ and $h\langle H_i\rangle^d$ respectively. Similarly,
we use $d_j$ and $e_j$ to denote products of the type
$f\langle\overline\Delta_j\rangle^u$ and
$f\langle\overline\Delta_j\rangle^d$ respectively. The MSSM vevs
are given by $v_u=X^*_{1\beta}\langle\phi_{u\beta}\rangle$ and
$v_d=Y^*_{1\beta}\langle\phi_{d\beta}\rangle$, where we use $v_u$
and $v_d$ to denote vevs of $H^{\rm MSSM}_u$ and $H^{\rm MSSM}_d$
respectively. The Yukawa couplings and vevs of Higgs fields in
general are complex, and there are 18 complex parameters. We
choose a basis in which the down-quark mass matrix is diagonalized
and set $b_3=0,b_4=0,$ and $b_5=0$. Note this is our main
difference with Ref.{\cite{s4LM}}, where they choose a basis in
which up-quark mass matrix is diagonal and set off-diagonal
entries of $M_u$ to zeros, which leads to small solar mixing
angle. In the basis we choose, the charged lepton mass matrix is
also diagnolized. Therefore, the phases of $b_0,b_1,b_2,e_0,e_1,$
and $e_2$ can be rotated away by redefining 3 right-handed
down-type quarks fields and three right-handed charged leptons. We
treat $b_0,b_1,b_2,e_0,e_1,$ and $e_2$ as real parameters in later
analysis, and they can be determined by the masses of down-quark
and charged lepton completely.

Because the mass matrix of down-quark sector is diagnolized and
$M_u$ is symmetric, one can have
\begin{eqnarray}\label{upmass2}
M_u=V_{CKM}^T\hat{M}_uV_{CKM},
\end{eqnarray}
where $\hat{M}_u\equiv diag(m_u,m_c,m_b)$. By fitting mass matrix of
up-quark in Eq.(\ref{upmass2}), parameters $a_3,a_4,a_5$ can be
determined. In addition, we get three conditions among the
parameters $a_0,a_1,a_2,d_0,d_1,$ and $d_2$. Therefore, there are
three complex parameters left to be determined by masses and mixings
of neutrino sector. Without loss of generality, we choose $d_0,d_1,$
and $d_2$ to be determined by fitting of neutrino sector. And Dirac
neutrino mass matrix can be written conveniently as
\begin{eqnarray}\label{numass}
M^D_\nu=V_{CKM}^T\hat{M}_u V_{CKM}-4m_t\left[\matrix{x&0&0\cr 0&
y&0\cr 0&0&z}\right]
\end{eqnarray}
with
\begin{eqnarray}\label{xdrelation}
 x\equiv \frac{1}{m_t}(d_0-2d_2),y\equiv
\frac{1}{m_t}(d_0+d_1+d_2),z\equiv \frac{1}{m_t}(d_0-d_1+d_2).
\end{eqnarray}
Because we know nothing about leptonic phases, in principle, there
is no constraint on the phases of $d_0,d_1,$ and $d_2$.

To see how this model can give a large atmospherical mixing angle,
we give an approximate analysis first. Using first order Wolfenstein
parameterization\cite{wolf} for the quark mixing,
$V_{CKM}^T\hat{M}_u V_{CKM}$ can be written as
\begin{eqnarray}\label{upmass}
m_t\left[\matrix{\lambda^6+A^2\lambda^6(1-i\eta-\rho)&\cdot&\cdot\cdot\cr
-\lambda^5-A^2\lambda^5(1-i\eta-\rho)&\lambda^4+A^2\lambda^4&\cdot\cdot\cdot\cr
A\lambda^3(1-i\eta-\rho)&-A\lambda^2&1}\right]
\end{eqnarray}
where we use $m_c/m_t\simeq\lambda^4$ and $m_u/m_t\simeq\lambda^8$.
Therefore, to get near maximal mixing of $\theta_{23}$, $y$ and $z$
should satisfy
\begin{eqnarray}
\lambda^4(1+A)-4y\simeq1-4z.
\end{eqnarray}

\section{Detailed Numerical Analysis}
 To see if the model is phenomenologically acceptable,
 we first fit the masses of the charged leptons and down-type quarks using
the mass values of leptons and quarks at the GUT scale with
$\tan\beta=10$\footnote{In this model, the value of $\tan\beta$ is
not determined as in MSSM. We take $\tan\beta=10$ as an example.}
given in Ref.\cite{dasparida}:
\begin{center}
\begin{tabular}{|c||c|}\hline\label{gutmass}
input observable & ${\rm tan\beta}=10$ \\ \hline $m_u$ (MeV) &
$0.7238^{+0.1365}_{-0.1467}$ \\\hline $m_c$ (MeV) &
$210.3273^{+19.0036}_{-21.2264}$
\\ \hline $m_t$ (GeV) & $82.4333^{+30.2676}_{-14.7686}$ \\\hline$m_d$ (MeV) &
$1.5036^{+0.4235}_{-0.2304}$
\\\hline
$m_s$ (MeV) & $29.9454^{+4.3001}_{-4.5444}$
\\\hline
$m_b$ (GeV) & $1.0636^{+0.1414}_{-0.0865}$
\\\hline
$m_e$ (MeV) & $0.3585^{+0.0003}_{-0.0003}$  \\\hline $m_{\mu}$ (MeV)
& $75.6715^{+0.0578}_{-0.0501}$ \\\hline $m_{\tau}$ (GeV) &
$1.2922^{+0.0013}_{-0.0012}$\\ \hline
\end{tabular}
\end{center}

We use standard parametrization form for the $V_{CKM}$ and take the
following values at the scale $\rm{Mz}$\cite{cpreview}:
$\sin\theta_{q12}=0.2272,\sin\theta_{q13}=0.00382,\sin\theta_{q23}=0.04178$
and the CP phase $\delta_q=\frac{\pi}{3}$, where we use subscript q
to distinguish them from the lepton section mixing angles. And we
use RGE running factor $\eta=0.8853$. At the GUT scale, we have the
$V_{CKM}$
\begin{eqnarray}
\left[
\begin{array}{ccc}
0.973841 & 0.227198 & 0.00169092-0.00292876 i\\
 -0.227079-0.000134603 i& 0.97298-0.000031403 i&0.0369876 \\
  0.00675837-0.00284968 i & -0.0364044-0.000664834 i& 0.99912\\
\end{array}
\right]
\end{eqnarray}

\subsection{Quark and charged lepton sector}
Using the central values of charged lepton and down-quark masses at
GUT scale, $b_0,b_1,b_2,e_0,e_1$ and $e_2$ are solved from
Eq.(\ref{lmass1}) and Eq.(\ref{dmass1}) (in Mev)
\begin{equation}
\begin{array}{rlrlrl}
b_0=387.756,~b_1=-539.649,~b_2=193.27,~
e_0=-22.7734,~e_1=22.8717,~e_2=-11.5298.
\end{array}
\end{equation}
For up-quark sector, by solving Eq.(\ref{upmass2}) and
Eq.(\ref{upmass1}), we get values of $a_3,a_4,a_5$ and three
conditions for $a_0,a_1,a_2,d_0,d_1,d_2$ (in Mev):
\begin{eqnarray}\label{upmasssolution}
\nonumber &&a_3=-2990.72-i54.757,~a_4=554.859-i234.705,~a_5=-66.748+i8.155,\\
\nonumber &&a_0-2a_2+d_0-2d_2=14.628-i3.162,~a_0+a_1+a_2+d_0+d_1+d_2=308.363+i3.977,\\
&&a_0-a_1+a_2+d_0-d_1+d_2=82288.5-i7.169\times10^{-6}.
\end{eqnarray}
We can see that accommodation of hierarchical structure of fermions
masses is realized by adjusting the parameters, $S4$ flavor symmetry
itself does not provide hints on it\footnote{One can use softly
broken discrete flavor symmetry to understand fermion mass hierachy,
see for example\cite{shafiz}.}.

\subsection{Neutrino sector}

In this model, the light neutrino mass matrix is given by type-I
seesaw\cite{seesaw}. The mass matrix of right-handed neutrinos is
proportional to an identity matrix due to the $S4$ quantum number
assignment, therefore the Dirac mass matrix $M^D$ determines the
lepton sector mixing because the charged lepton mass matrix is
diagnolized.
\begin{eqnarray}
M_\nu=-\frac{1}{f_0v_R}{M^D_\nu}^T M^D_\nu.
\end{eqnarray}
This model gives hierarchical neutrino mass spectrum naturally. One
can choose $f_0\sim1$ and $v_R\sim 10^{14}\rm{GeV}$, so the mass of
the heaviest light neutrino is around $10^{-2}-10^{-1}\rm{eV}$.

The fit of neutrino sector are found by scanning whole parameter
space spanned by $x,y$ and $z$ under the constrain of the current
experiment requirements.

We choose the standard parametrization for the lepton sector mixing:
\begin{eqnarray}
U=\left[
 \begin{array}{ccc}
 c_{12}c_{13} & s_{12}c_{13} & s_{13}e^{-i\delta}\\
 -c_{23}s_{12}-s_{23}s_{13}c_{12}e^{i\delta} &
 c_{23}c_{12}-s_{23}s_{13}s_{12}e^{i\delta} & s_{23}c_{13}\\
 s_{23}s_{12}-c_{23}s_{13}c_{12}e^{i\delta} &
 -s_{23}c_{12}-c_{23}s_{13}s_{12}e^{i\delta} & c_{23}c_{13}
 \end{array}
 \right].{\rm diag}(e^{-i\varphi_1/2},e^{-i\varphi_2/2},1)
\end{eqnarray}

with $c_{ij}\equiv \cos\theta_{ij}$ and
$s_{ij}\equiv\sin\theta_{ij}$. $\delta$ is the Dirac phase and
$\varphi_1,\varphi_2$ are Majorona phases of neutrinos. These phases
have range from $0$ to $2\pi$.

We take $3\sigma$ experiment bound\cite{valle}:
\begin{eqnarray}\label{bound}
\nonumber &&~~~~~~0.24\leq\sin^2\theta_{12}\leq0.40\\
\nonumber &&~~~~~~0.34\leq\sin^2\theta_{23}\leq0.68\\
\nonumber &&~~~~~~~~~~~\sin^2\theta_{13}\leq0.040\\
&& 0.024\leq\Delta m^2_{\odot}/\Delta m^2_{ATM}\leq 0.040.
\end{eqnarray}

As mentioned earlier $x,y,$ and $z$ generally are complex numbers.
For the most general CP-violation case, we treat the phases of
$x,y,$ and $z$ as random input numbers with range $0-2\pi$. The
results are shown in Fig.(1). In this case, $\sin\theta_{13}$ has
wide range from zero to the current bound with the most probable
values $0.02-0.09$ as shown in Fig.1(a). Fig.1(b) shows the
correlation between $\sin\theta_{23}$ and $\sin\theta_{13}$.
Fig.1(c) is the value distribution of Dirac CP-violation phase in
the lepton sector. The allowed range of $\delta$ is quite large from
$0$ to $2\pi$ radians with the most probable values $2-4$ radians.
Two Majorana phases $\varphi_1$ and $\varphi_2$ have wide range from
$0$ to $2\pi$ as shown in Fig.1(d), which is expected.

Now we consider an interesting special case where $x,y,$ and $z$ are
all real. Note the complexity of $f_0v_R$ only contributes an
overall phase to the light neutrino mass matrix, which can be
rotated away. Therefore, in this case leptonic CP-violation phases
have same source as CKM phase.
\begin{figure}[t!] \centerline{
\includegraphics[scale=0.7]{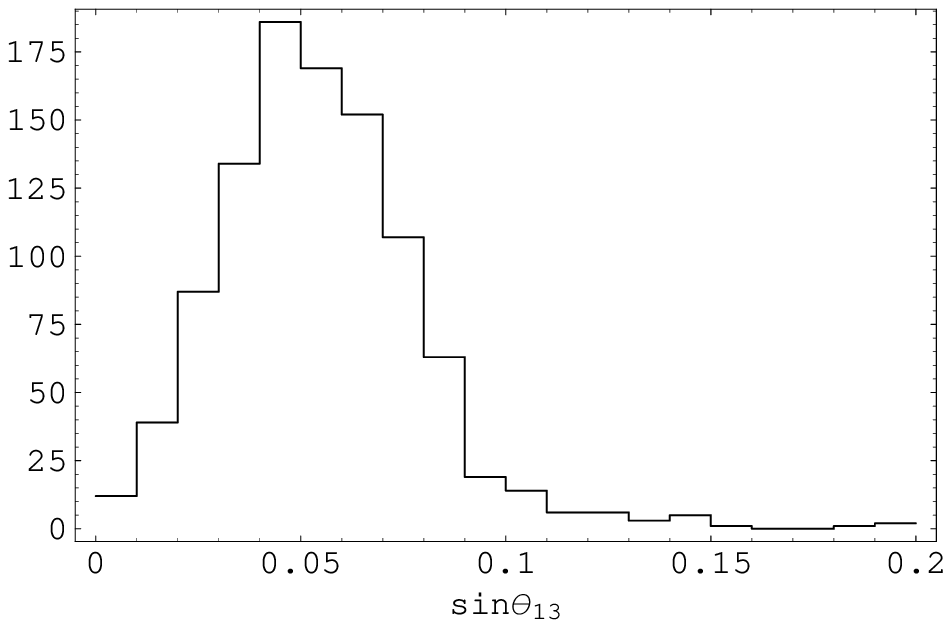}
\includegraphics[scale=0.7]{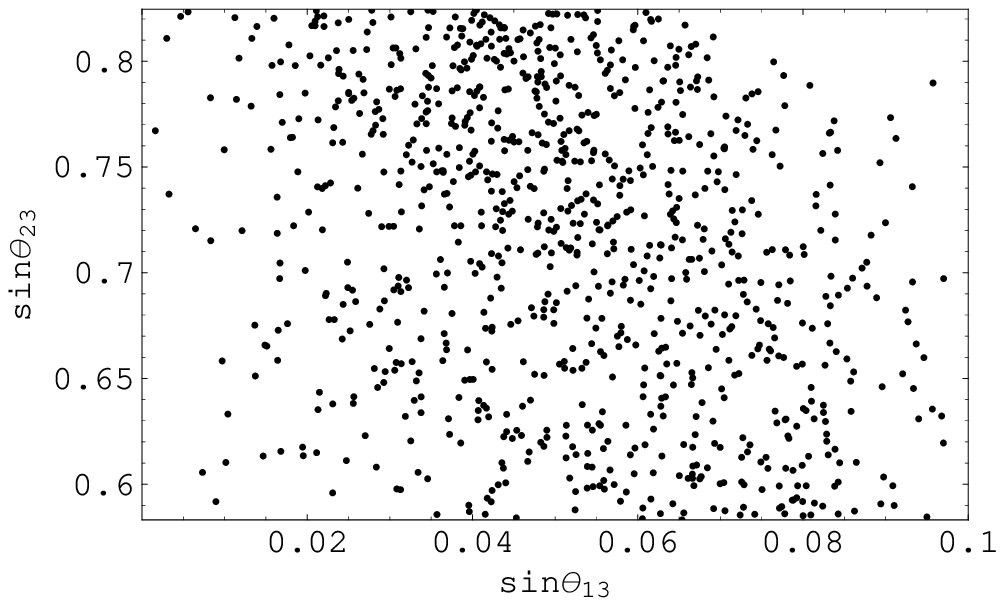}}
\vspace*{8mm}\centerline{
\includegraphics[scale=0.7]{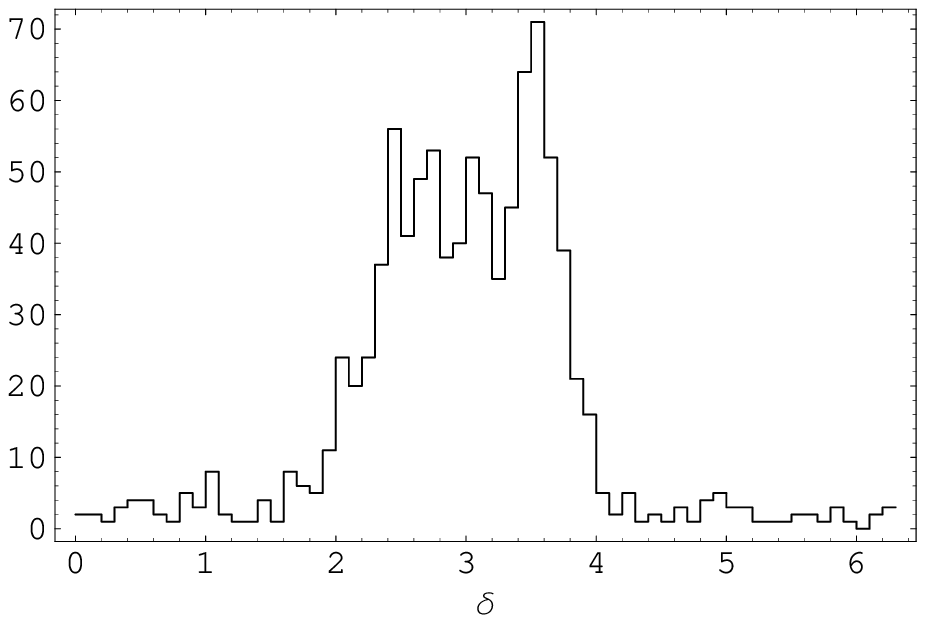}
\includegraphics[scale=0.7]{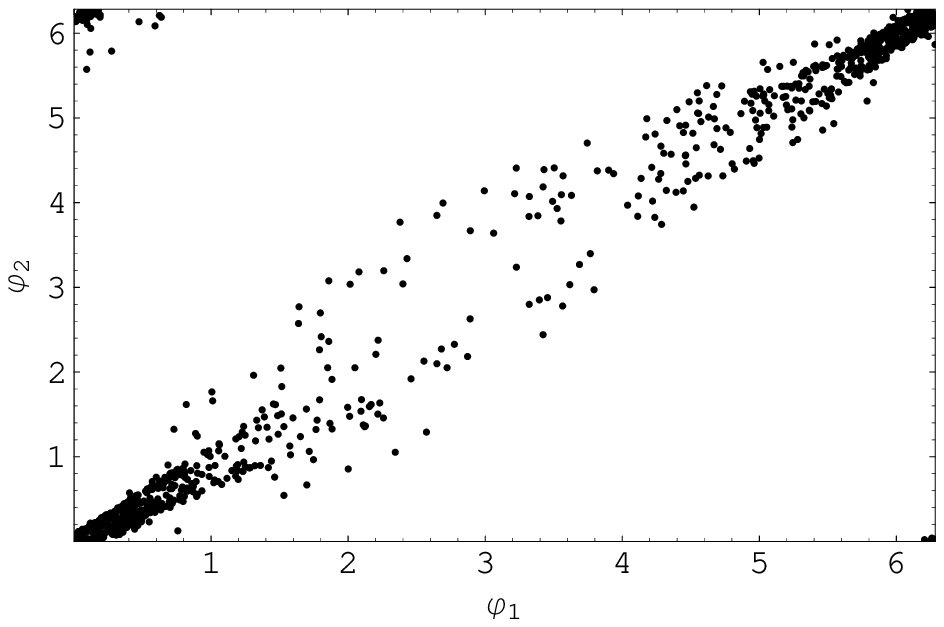}
} \caption{Numerical analysis for the most general case where $x,y,$
and $z$ are complex consistent with current experimental bound
Eq.(\ref{bound}). (a)~Value distribution of $\sin\theta_{13}$.
(b)~Correlation between $\sin\theta_{23}$ and $\sin\theta_{13}$.
(c)~Value distribution of leptonic Dirac CP-violation phase.
(d)~Scatter plot of two Majorana CP-violation phases $\varphi_1$ and
$\varphi_2$.}
\begin{picture}(0,0)
\put(-105,260){(a)} \put(105,260){(b)} \put(-105,110){(c)}
\put(105,110){(d)}
\end{picture}
\end{figure}

\begin{figure}[t!]
\centerline{
\includegraphics[scale=0.7]{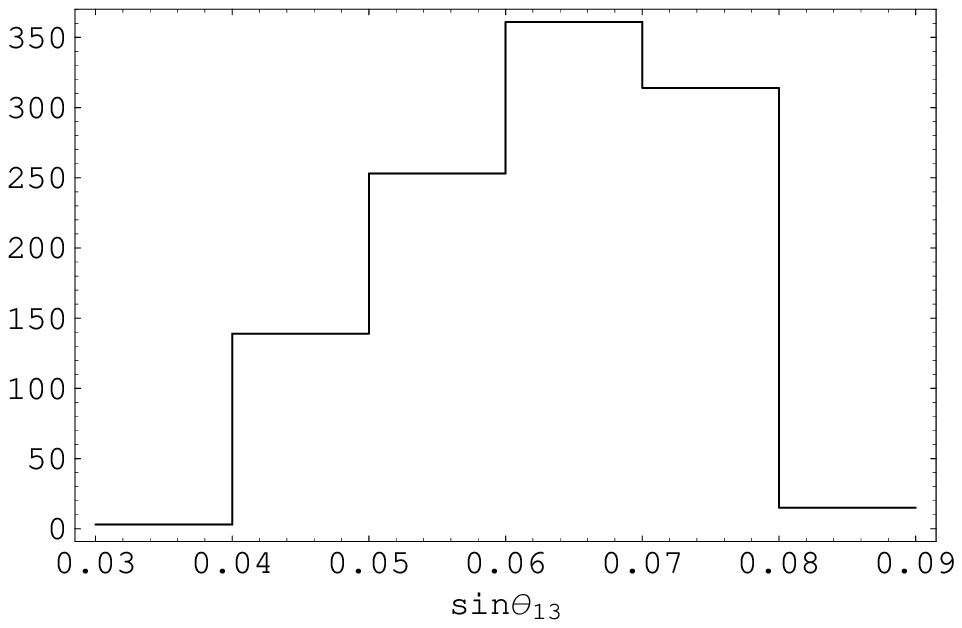}
\includegraphics[scale=0.7]{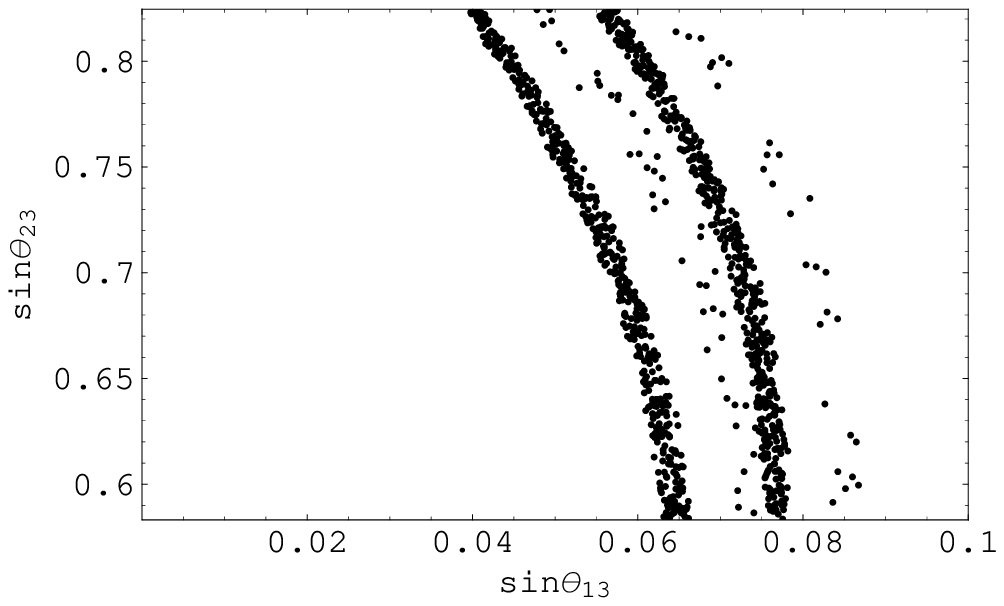}
} \vspace*{8mm} \centerline{
\includegraphics[scale=0.7]{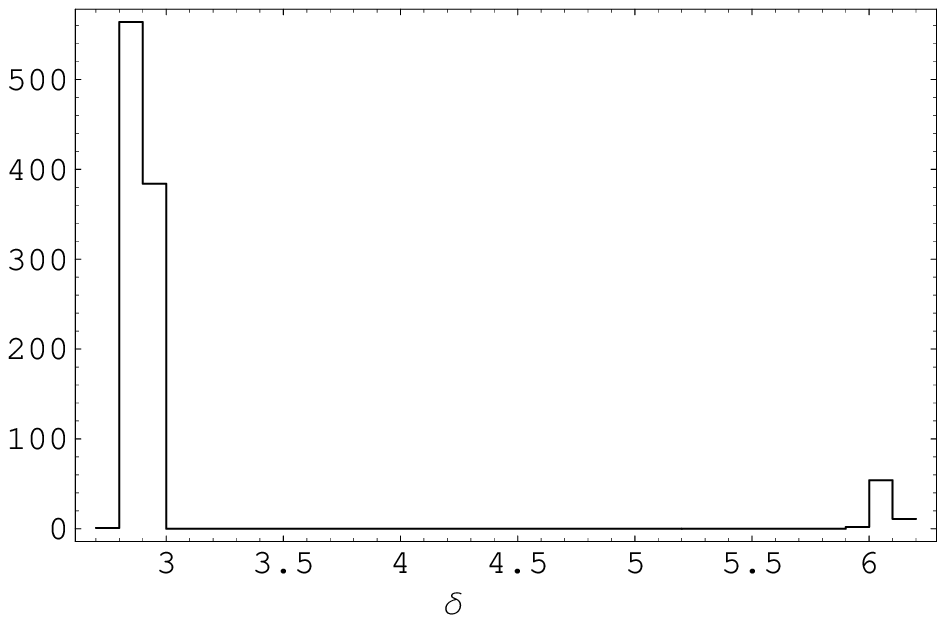}
\includegraphics[scale=0.7]{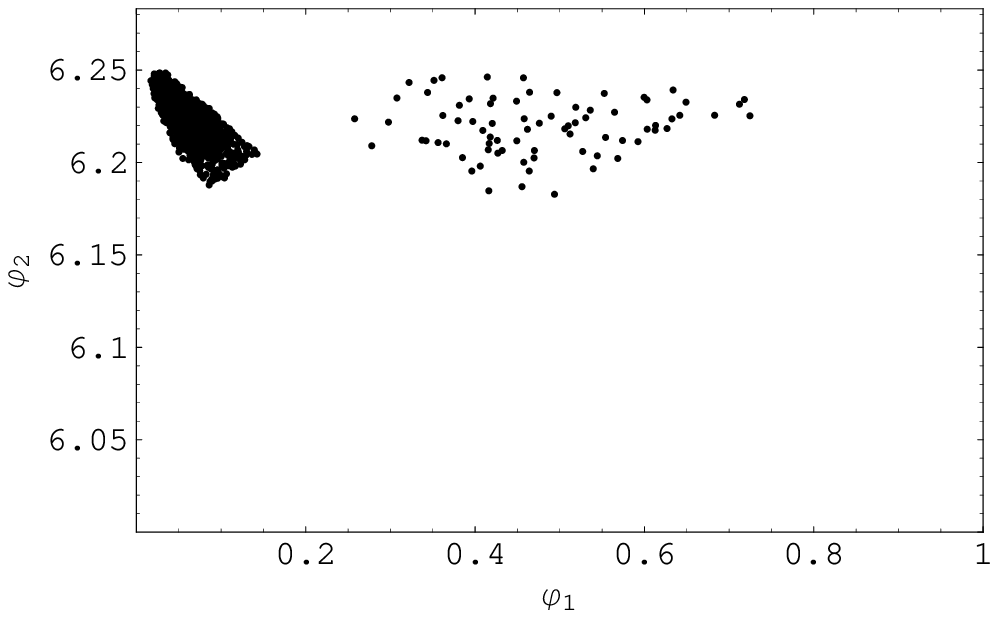}
} \caption{Numerical analysis for case where $x,y,z$ are real
consistent with current experimental bound Eq.(\ref{bound}).
(a)~Value distribution of $\sin\theta_{13}$. (b)~Correlation between
$\sin\theta_{23}$ and $\sin\theta_{13}$. (c)~Value distribution of
leptonic Dirac CP-violation phase. (d)~Scatter plot of two Majorana
CP-violation phases $\varphi_1$ and $\varphi_2$.}
\begin{picture}(0,0)
\put(-105,260){(a)} \put(105,260){(b)} \put(-110,110){(c)}
\put(110,110){(d)}
\end{picture}
\end{figure}
The allowed range $0.03-0.09$ for $\sin\theta_{13}$ is narrower
compared to the general case, and the most probable range is
$0.04-0.08$ as shown in Fig.2(a). Unlike Fig.1(b), Fig.2(b) exhibits
an interesting correlation between $\sin\theta_{23}$ and
$\sin\theta_{13}$. If we take the central value of
$\theta_{23}=\frac{\pi}{4}$, we can get two much narrower ranges for
$\sin\theta_{13}$. One is $0.055-0.06$, and the other is
$0.070-0.075$. The values of $\delta$ are $2.8-3$ radians, and
$6.0-6.1$ with small possibility as shown in Fig.2(c). Fig.2(d)
shows the allowed values of two Majorana phases. Note this parameter
region is just left-up corner of Fig.1(d) for the most general case.
The most probable value ranges for $\varphi_1$ and $\varphi_2$ are
$0.02-0.15$ radians and $6.19-6.25$ radians respectively.

For illustration, we give a typical example of fit for this case. We
take
\begin{eqnarray}
x=0.0139726,~y=0.025914,~z=0.273173
\end{eqnarray}
and solve $d_0,d_1,d_2,a_0,a_1,a_2$ from Eq.(\ref{xdrelation}) and
Eq.(\ref{upmasssolution})(in Mev)
\begin{eqnarray}
\nonumber
&&d_0=8602.18,~d_1=-10191.2,~d_2=3725.19,~a_0=18935+i0.271681,
\\&&a_1=-30798.9+i1.9887,~a_2=10036.1+i1.71701.
\end{eqnarray}

With these parameters values as input, one then obtains for the
neutrino parameters
\begin{equation}
\begin{array}{rl}
\sin\theta_{12}\simeq0.53,&~\sin\theta_{23}\simeq0.73\\
\sin\theta_{13}\simeq0.054,&\Delta m^2_{\odot}/\Delta
m^2_{ATM}\simeq0.031.
\end{array}
\end{equation}
And light neutrino masses are $m_1=0.00774{\rm eV},~m_2=0.0118{\rm
eV},~m_3=0.051{\rm eV}$, which are normalized by $\Delta
m_{31}^2=2.6\times10^{-3}{\rm eV}$. The Dirac phase appearing in MNS
matrix is $\delta=2.84$ radians. And two Majorona phases are (in
radians): $\varphi_1=0.093,~\varphi_2=6.21$. The Jarlskog
invariant\cite{jarlskog} has the value $J{\rm
cp}=1.80\times10^{-3}$. One can evaluate the effective neutrino mass
for the neutrinoless double beta decays process to be
\begin{eqnarray}
\nonumber |\sum U^2_{ei}m_{\nu i}|\simeq0.009~{\rm eV}.
\end{eqnarray}

\section{Summary and conclusion}
In summary, we build a supersymmetric $SO(10)$ model with $S4$
flavor symmetry. The three dimensional irreducible representation
of $S4$ group unify three generations of fermions horizontally.
${\bf 10}$ and ${\bf \overline{126}}$ Higgs fields have been used
to give the Yukawa couplings and generate all the masses and
mixings of quarks and leptons. This model accommodates all
obervables including CKM CP-Violation phase. We studied the
prediction of this model in the neutrino sector. For the most
general CP-violation case, this model gives the most probable
values $0.02-0.09$ for $\sin\theta_{13}$. In a special case where
leptonic phases have same CP-violation source as CKM phase, one
gets narrower range $0.03-0.09$ for $\sin\theta_{13}$ with the
most probable values $0.04-0.08$.

In the model we present here, the masses of light neutrinos purely
come from the type-I seesaw\cite{seesaw}. Generally, one also can
include the contribution from type-II seesaw\cite{seesaw2}, which
can generate a scenario with degenerate neutrino mass spectrum
naturally because of the $S4$ symmetry if the type-II seesaw
dominates the contribution to the light neutrinos masses. It is
interesting to study the mixing pattern and its radioactive
stability. We leave this possibility for future work.

\section{Acknowledgments}
The authors thank R.N. Mohapatra for suggesting the problem and
useful discussion, K. Hsieh, J.H. Kang and P. Rastogi for comments.
H.B.Y. specially thanks C. Hagedorn for helpful comments on the
discrete symmetry. This work is supported by the National Science
Foundation Grant No. PHY-0354401.


\begin{thebibliography}{99}

\bibitem{rev} For a recent review of the theoretical issues in
neutrino physics, see R. N. Mohapatra and A. Y. Smirnov,
arXiv:hep-ph/0603118, to appear in Ann. Rev. Nucl. Science, {\bf 56}
(2006).

\bibitem{mutau} T. Fukuyama and H. Nishiura, hep-ph/9702253;
R. N. Mohapatra and S. Nussinov, Phys. Rev. {\bf D 60}, 013002
(1999); E. Ma and M. Raidal, Phys. Rev. Lett. {\bf 87}, 011802
(2001); C. S. Lam, hep-ph/0104116; T. Kitabayashi and M. Yasue,
Phys.Rev. {\bf D67} 015006 (2003); W. Grimus and L. Lavoura,
hep-ph/0305046; 0309050; Y. Koide, Phys.Rev. {\bf D69}, 093001
(2004);Y. H. Ahn, Sin Kyu Kang, C. S. Kim, Jake Lee, hep-ph/0602160;
 A. Ghosal, hep-ph/0304090; for examples of such theories,
see W. Grimus and L. Lavoura, Phys.\ Lett.\ B {\bf 572}, 189 (2003);
W.~Grimus and L.~Lavoura, J.\ Phys.\ G {\bf 30}, 73 (2004).

\bibitem{moh}  W. Grimus, A. S.Joshipura, S. Kaneko, L.
Lavoura, H. Sawanaka, M. Tanimoto, hep-ph/0408123; R. N. Mohapatra,
JHEP, {\bf 0410}, 027 (2004); A. de Gouvea, Phys.Rev. {\bf D69},
093007 (2004); R. N. Mohapatra and W. Rodejohann, Phys. Rev. {\bf D
72}, 053001 (2005); T. Kitabayashi and M. Yasue, Phys. Lett,. {\bf B
621}, 133 (2005);  R.~N.~Mohapatra and S.~Nasri,
  Phys.\ Rev.\ D {\bf 71}, 033001 (2005); R.~N.~Mohapatra, S.~Nasri and H.~B.~Yu,
  Phys.\ Lett.\ B {\bf 615}, 231 (2005); R.~N.~Mohapatra, S.~Nasri and H.~B.~Yu,
  Phys.\ Rev.\ D {\bf 72}, 033007 (2005); Z.~z.~Xing,
  arXiv:hep-ph/0605219; Z.~z.~Xing, H.~Zhang and S.~Zhou, arXiv:hep-ph/0607091.

\bibitem{mutaugut}  K. Matsuda and H. Nishiura,
 Phys.\ Rev.\ D {\bf 73}, 013008 (2006);  A. Joshipura, hep-ph/0512252;
R. N. Mohapatra, S. Nasri and H.~B.~Yu, Phys. Lett. {\bf B 636}, 114
(2006).
\bibitem{s3}L.~J.~Hall and H.~Murayama, Phys.\
  Rev.\ Lett.\  {\bf 75}, 3985 (1995); C.~D.~Carone, L.~J.~Hall and H.~Murayama, Phys.\ Rev.\ D
{\bf 53}, 6282 (1996);
  P.~F.~Harrison and W.~G.~Scott, Phys.\ Lett.\ B {\bf 557}, 76
  (2003);  J.~Kubo, A.~Mondragon, M.~Mondragon and
  E.~Rodriguez-Jauregui, Prog.\ Theor.\ Phys.\  {\bf 109}, 795 (2003)
  Erratum-ibid.\  {\bf 114}, 287 (2005); S.~L.~Chen, M.~Frigerio and
  E.~Ma, Phys.\ Rev.\ D {\bf 70}, 073008 (2004)
  Erratum-ibid.\ D {\bf 70}, 079905 (2004); L.~Lavoura and E.~Ma, Mod.\ Phys.\ Lett.\ A {\bf 20}, 1217
  (2005); R.~Dermisek and
  S.~Raby, Phys.\ Lett.\ B {\bf 622}, 327 (2005);
  F.~Caravaglios and S.~Morisi, arXiv:hep-ph/0503234;
  W.~Grimus and L.~Lavoura, JHEP {\bf 0508}, 013 (2005);  N.~Haba and K.~Yoshioka,
  Nucl.\ Phys.\ B {\bf 739}, 254 (2006); S.~Morisi, arXiv:hep-ph/0604106;   R.~N.~Mohapatra, S.~Nasri and H.~B.~Yu,
  Phys.\ Lett.\ B {\bf 639}, 318 (2006); R.~Jora, S.~Nasri and J.~Schechter,
hep-ph/0605069.
  S.~Morisi, arXiv:hep-ph/0605167.

\bibitem{s4}   S.~Pakvasa and H.~Sugawara, Phys.\ Lett.\ B {\bf 82},
  105 (1979); E. ~Derman and H.-S. ~Tsao, Phys. \ Rev. \ D  {\bf
  20}, 1207 (1979); D.-G. ~Lee and R. ~N. ~Mohapatra, Phys. \ Lett. \ B {\bf 329}, 463
(1994);  R.~N.~Mohapatra, M.~K.~Parida and G.~Rajasekaran, Phys.\
Rev.\ D {\bf 69}, 053007 (2004) ;   E.~Ma, Phys.\ Lett.\ B {\bf
632}, 352 (2006);   C.~Hagedorn, M.~Lindner and R.~N.~Mohapatra,
JHEP {\bf 0606}, 042(2006).

\bibitem{a4}E.~Ma and G.~Rajasekaran, Phys.\ Rev.\ D {\bf 64}, 113012 (2001); K.~S.~Babu, E.~Ma and J.~W.~F.~Valle,
  Phys.\ Lett.\ B {\bf 552}, 207 (2003);  K.~S.~Babu and X.~G.~He, arXiv:hep-ph/0507217; A. Zee, Phys. Lett. B 630, 58 (2005);  S.~L.~Chen, M.~Frigerio and E.~Ma,
  Nucl.\ Phys.\ B {\bf 724}, 423 (2005); G.~Altarelli and F.~Feruglio, Nucl.\ Phys.\ B {\bf 741}, 215 (2006); X.~G.~He, Y.~Y.~Keum and R.~R.~Volkas,
  JHEP {\bf 0604}, 039 (2006); B.~Adhikary, B.~Brahmachari, A.~Ghosal, E.~Ma and M.~K.~Parida,
  Phys.\ Lett.\ B {\bf 638}, 345 (2006); E.~Ma, H.~Sawanaka and M.~Tanimoto, arXiv:hep-ph/0606103; E.~Ma,
  arXiv:hep-ph/0607190.




\bibitem{d4} W.~Grimus and L.~Lavoura, Phys.\ Lett.\ B {\bf
    572}, 189 (2003);   W.~Grimus, A.~S.~Joshipura, S.~Kaneko,
  L.~Lavoura and M.~Tanimoto, JHEP {\bf 0407}, 078 (2004); G.~Seidl,
 arXiv:hep-ph/0301044;   T.~Kobayashi, S.~Raby and R.~J.~Zhang, Nucl.\ Phys.\ B {\bf 704}, 3 (2005).
\bibitem{d5}  E.~Ma, arXiv:hep-ph/0409288; C.~Hagedorn, M.~Lindner and F.~Plentinger,Phys.\ Rev.\ D {\bf 74}, 025007 (2006)


\bibitem{s4HLM}  C.~Hagedorn, M.~Lindner and R.~N.~Mohapatra, JHEP {\bf 0606},
042(2006).
\bibitem{lepto}  J.~R.~Ellis, M.~Raidal and T.~Yanagida, Phys.\ Lett.\ B {\bf 546}, 228
(2002);  A.~Pilaftsis and T.~E.~J.~Underwood, Nucl.\ Phys.\ B {\bf
692}, 303 (2004).
\bibitem{s4LM}  D.~G.~Lee and R.~N.~Mohapatra, Phys.\ Lett.\ B {\bf 329}, 463 (1994)



\bibitem{minso10}  K.~S.~Babu and R.~N.~Mohapatra,
   Phys.\ Rev.\ Lett.\  {\bf 70}, 2845 (1993).

\bibitem{others} T.~E.~Clark, T.~K.~Kuo and N.~Nakagawa,  Phys.\ Lett.\ B
{\bf 115}, 26 (1982);  C.~S.~Aulakh and  R.~N.~Mohapatra, Phys.\ Rev.\ D
{\bf 28}, 217 (1983); C.~S.~Aulakh, B.~Bajc, A.~Melfo, G.~Senjanovic and
F.~Vissani,  Phys.\ Lett.\ B {\bf 588}, 196 (2004);  B.~Bajc, A.~Melfo,
G.~Senjanovic and F.~Vissani,  Phys.\ Rev.\ D {\bf 70}, 035007 (2004);
T.~Fukuyama, A.~Ilakovac, T.~Kikuchi, S.~Meljanac and N.~Okada,
  Phys.\ Rev.\ D {\bf 72}, 051701 (2005); H.~S.~Goh, R.~N.~Mohapatra and
S.~Nasri,  Phys.\ Rev.\ D {\bf 70}, 075022 (2004).

\bibitem{kubo1}  T.~Kobayashi, J.~Kubo and H.~Terao, Phys.\ Lett.\ B {\bf 568}, 83
(2003).
\bibitem{kubo2}   K.~Y.~Choi, Y.~Kajiyama, H.~M.~Lee and J.~Kubo,
  Phys.\ Rev.\ D {\bf 70}, 055004 (2004).
\bibitem{wolf}  L.~Wolfenstein, Phys.\ Rev.\ Lett.\  {\bf 51}, 1945 (1983).
\bibitem{dasparida}  C.~R.~Das and M.~K.~Parida,
  Eur.\ Phys.\ J.\ C {\bf 20}, 121 (2001).
\bibitem{cpreview}
  A.~Hocker and Z.~Ligeti, arXiv:hep-ph/0605217.
\bibitem{shafiz} Q.~Shafi and Z.~Tavartkiladze, Phys.\ Lett.\ B {\bf 594}, 177 (2004).
\bibitem{seesaw}P. Minkowski, Phys. Lett. {\bf B67 }, 421
(1977); M.~Gell-Mann, P.~Ramond, and R.~Slansky, {\it Supergravity}
(P.~van Nieuwenhuizen et al. eds.), North Holland, Amsterdam, 1980,
; T.~Yanagida, in {\it Proceedings of the Workshop on the Unified
Theory and the Baryon Number in the Universe} (O.~Sawada and
A.~Sugamoto, eds.), KEK, Tsukuba, Japan, 1979; S.~L. Glashow, {\it
The future of elementary particle physics}, in {\it Proceedings of
the 1979 Carg{\`e}se Summer Institute on Quarks and Leptons}
(M.~L{\'e}vy et al. eds.), Plenum Press, New York, 1980, pp.~687;
R.~N. Mohapatra and G.~Senjanovi{\'c}, Phys. Rev. Lett. \textbf{44}
912 (1980).
\bibitem{seesaw2}  G. Lazarides, Q. Shafi and C. Wetterich,
Nucl.Phys.{\bf B181}, 287 (1981); R. N. Mohapatra and G.
Senjanovi\'c, Phys. Rev. {\bf D 23}, 165 (1981).
\bibitem{valle}   M.~Maltoni, T.~Schwetz, M.~A.~Tortola and J.~W.~F.~Valle,
   New J.\ Phys.\  {\bf 6}, 122 (2004). See Appendix C of
   aiXiv:hep-ph/0405172 for updated fit.
\bibitem{jarlskog} C. Jarlskog, Phys. Rev. Lett. {\bf 55}, 1039
(1985).
 \end{thebibliography}
\end{document}